\documentclass[paper=a4, fontsize=11pt]{scrartcl} 

\usepackage[T1]{fontenc} 
\usepackage[utf8]{inputenc}
\usepackage[english]{babel} 
\usepackage{amsmath,amsfonts,amsthm,amssymb} 

\usepackage{sectsty} 
\allsectionsfont{\centering \normalfont\scshape} 
\usepackage{placeins}
\usepackage{indentfirst}
\usepackage{fancyhdr} 
\usepackage{adjustbox}
\usepackage{authblk}
\usepackage{pifont}

\numberwithin{equation}{section}
\numberwithin{figure}{section}
\numberwithin{table}{section}

\newcommand*\pFqskip{8mu}
\catcode`,\active
\newcommand*\pFq{\begingroup
        \catcode`\,\active
        \def ,{\mskip\pFqskip\relax}%
        \dopFq
}
\catcode`\,12
\def\dopFq#1#2#3#4#5{%
        {}_{#1}F_{#2}\biggl[\genfrac..{0pt}{}{#3}{#4};#5\biggr]%
        \endgroup
}

\title{Analyticity of the free energy for~quantum Airy structures}

\author{Błażej Ruba}
\affil{\textit{Institute of Physics, Jagiellonian University}}
\affil{\textit{prof. Łojasiewicza 11, 30-348 Kraków, Poland}}

\date{\today}

\begin{document}

\maketitle

\abstract{It is shown that the free energy associated to a finite dimensional Airy structure is an analytic function at each finite order of the $\hbar$ expansion. Semiclassical series itself is in general divergent. Calculations are facilitated by putting the topological recursion equations into a form exhibiting more explicitly the semiclassical geometry. This formulation involves certain differential operators on the characteristic variety, which are found to satisfy a~Lie algebra cocycle condition. It is proven that this cocycle is a coboundary. Developed formalism is applied in specific examples. In the case of a divergent $\hbar$ series, a simple resummation is performed. Analytic properties of the obtained partition functions are investigated.}


\section{Introduction and results}

Airy structures were introduced in \cite{Kontsevich} as a reformulation and generalization of the topological recursion \cite{EyOr} for spectral curves. They encode data necessary to formulate certain recursive equations encountered in matrix models \cite{MatrixEynard, MatrixBorot, TR_CFT_CHJMS}, conformal field theory \cite{TR_CFT_Kostov}, enumerative geometry \cite{Alexandrov_Hurwitz, Bouchard_Hurwitz}, and other applications (see \cite{Eynard_TR_Review, Borot} for pedagogical reviews) as differential equations. This approach makes manifest symmetry properties of the solutions and strong symplectic flavour of the problem. Further study of Airy structures was undertaken in \cite{ABCD, W_algebras, SuperAiry}.


Recall that an Airy structure is a Lie algebra of differential operators
\begin{subequations}
\begin{gather}
L_i = \hbar \partial_i - \frac{1}{2} A_{ijk} x^j x^k - \hbar B_{ij}^k x^j \partial_k - \frac{\hbar^2}{2} C_i^{jk} \partial_j \partial_k - \hbar D_i, \label{eq:Li_definition} \\
[L_i, L_j] = \hbar f_{ij}^k L_k, \label{eq:Li_bracket}
\end{gather}
\end{subequations}
where indices $i, j, k \in \{ 1,..., n \}$. Repeated indices are summed over in all equations. In general infinite index sets are allowed, but this case will not be considered here. Symbols $A,B,C,D$ and $f$ are tensors obeying a number of polynomial equations following from (\ref{eq:Li_bracket}), whose exact form will not be important here. All calculations are performed over the field $k = \mathbb R$ or $\mathbb C$.


\vspace{1em}

It was shown in \cite{Kontsevich} that to every quantum Airy structure one may associate a unique free energy, i.e.\ a formal series $F$ in $\hbar$ and $x^i$ such that $F(\hbar,0)=0$, $\partial_i F(0,0) = \partial_i \partial_j F(0,0)=0$ and
\begin{equation}
L_i e^{\frac{F}{\hbar}}=0. \label{eq:Z_annihilated}
\end{equation}
Due to the presence of a negative power of $\hbar$ in the exponent, it is not immediately obvious that the partition function $Z= e^{\frac{F}{\hbar}}$ may be made sense of as a formal series. However, due to conditions imposed on $F$, one may obtain a series involving only positive powers of indeterminates by performing a change of variables $x = \hbar^{\frac{1}{2}} x'$. Therefore all operations defined for formal power series, such as taking the exponential of $\frac{F}{\hbar}$ or inverting $Z$, can be applied without encountering meaningless expressions. In \cite{Kontsevich} a slightly different approach was proposed: the map $L_i \mapsto Z^{-1} L_i Z$ was regarded as an automorphism of the completed Weyl algebra, and equation (\ref{eq:Z_annihilated}) was reinterpreted as $(Z^{-1} L_i Z) \cdot 1=0$. Using the calculus of formal series one may show that $(Z^{-1} L_i Z) \cdot 1= Z^{-1} (L_i Z)$, so the two approaches are equivalent.


\vspace{1em}

By construction, the free energy may be expanded in powers of $\hbar$,
\begin{equation}
F(\hbar, x) = \sum_{g =0}^{\infty} \hbar^g F_g(x),
\label{eq:F_hbar_expansion}
\end{equation}
with $F_g \in k [[x]]$. One of the results of the present work is that there exists a neighbourhood $U$ of $0 \in k^n$ such that the series $F_g(x)$ converges for each $x \in U$ and $g \in \mathbb N$. The series (\ref{eq:F_hbar_expansion}) itself diverges in general, which will be demonstrated by an explicit example. Employed techniques are inspired by the lecture notes \cite{GofQ}.

\vspace{1em}

In the process of proving the claim, partition function is rewritten as
\begin{equation}
Z = e^{\frac{F_0(x)}{\hbar}} \rho_0(x) \left( 1+ \sum_{m=1}^{\infty} \hbar^m \psi_m(x) \right).
\label{eq:Z_our_ansatz}
\end{equation}
Terms of this expansion are interpreted as geometric objects on the Lagrangian submanifold $\Sigma_0$, locally given by $ y_i = \partial_i F_0(x) $, in the phase space $k^{2n}$ equipped with the symplectic form $\omega= dy_i \wedge dx^i$. Simply connected Lie group $G$ generated by the Lie algebra $\mathfrak g$ spanned by $L_i$ acts on $k^{2n}$ by hamiltonian transformations. Flow lines of this action are tangent to $\Sigma_0$. It is argued that $\rho_0$ is best thought of as a local expression for a half-form on $\Sigma_0$ of a charge\footnote{Charge $\delta_i$ is a Lie algebra cocycle. Its vanishing defines the Weyl quantization prescription.} $\delta_i = D_i - \frac{1}{2} B_{ij}^j$ with respect to the $G$-action. Weyl quantization scheme is distinguished by the condition $\delta_i=0$.  Functions $\psi_m$ may be interpreted as quantum corrections to this half-density. They satisfy a ladder of inhomogeneous transport equations
\begin{equation}
\xi_i \psi_m = \Delta_i \psi_{m-1},
\label{eq:xi_psi_Delta}
\end{equation}
where $\xi_i$ is the hamiltonian vector field corresponding to the generator $L_i$, restricted to $\Sigma_0$. $\Delta_i$ are commuting second order differential operators, built of the $C$ tensor and $\rho_0$. They obey a Lie algebra cocycle condition
\begin{equation}
[\xi_i, \Delta_j] - [\xi_j, \Delta_i] = f_{ij}^k \Delta_k.
\label{eq:Delta_cocycle}
\end{equation}

It is proven that it is possible to trivialize this cocycle, i.e.\ to find a second order differential operator $\kappa$ on a neighbourhood of zero in $\Sigma_0$ such that 
\begin{equation}
\Delta_i = [\xi_i, \kappa].
\end{equation}

\vspace{1em}

Incidentally, it follows from the presented proof that commutativity of $\Delta_i$ and the cocycle condition (\ref{eq:Delta_cocycle}) are the only conditions needed for existence, uniqueness and regularity of the partition function. More precisely, let $G$ be a~Lie group and let $\xi_i$ be a basis of its Lie algebra of (say, left-invariant) vector fields. Suppose that $U$ is a connected neighbourhood of the neutral element $0 \in G$, with $H^1_{\mathrm{dR}}(U)=0$. Let $\Delta_i$ be commuting differential operators on $U$ satisfying (\ref{eq:Delta_cocycle}). Then operators $T_i = \xi_i - \hbar \Delta_i$ satisfy $[T_i,T_j]= f_{ij}^k T_k$. Moreover, there exists a unique $\psi = 1 + \sum_{m=1}^{\infty} \hbar^m \psi_m \in C^{\infty}(U)[[\hbar ]]$ annihilated by all $T_i$. If $G$ is complex and $\Delta_i$ are holomorphic, then so is each $\psi_m$. Every Airy structure determines data of this type, but not every commuting cocycle $\Delta_i$ may be derived from an Airy structure.

\vspace{1em}

The whole formalism is applied in several examples. These illustrate that the system (\ref{eq:xi_psi_Delta}) provides an efficient tool for solving for $Z$, provided that the closed form of $F_0$ and $\rho_0$ may be found. This may be a challenging task by itself. Nevertheless, it is much easier than finding $Z$, mainly because the relevant differential equations are of the first (rather than second) order. Global analytic structure of the partition function on $\Sigma_0$ was investigated. In the examples considered, it was possible to extend the quantum corrections to meromorphic functions on $\Sigma_0$, possibly with poles on the ramification locus of the projection to the $x$ axis. In one case, the partition function turned out to be a divergent series in $\hbar$. It was possible to resum it into an expression with an essential singularity on a codimension one subvariety of $\Sigma_0$.


\section{Proofs}

It is sufficient to consider the case $k = \mathbb C$. Analyticity of $F$ in the real case then follows, because every Airy structure over $\mathbb R$ may be complexified.

\vspace{1em}

\textit{Classical Airy structure}

\vspace{1em}

To any given quantum Airy structure one may associate its classical limit, i.e.\ the set of hamiltonians
\begin{equation}
L_i^{\mathrm{cl}}(x,y) = y_i - \frac{1}{2} A_{ijk} x^j x^k - B_{ij}^k x^j y_k - \frac{1}{2} C_{i}^{jk} y_j y_k.
\end{equation}
They satisfy bracket relations $\{ L_i^{\mathrm{cl}}, L_j^{\mathrm{cl}} \} = f_{ij}^k L_k^{\mathrm{cl}}$, where Poisson bracket $\{ - , - \}$ on $k[x,y]$ is determined by the elementary relation $\{ y_i, x^j \} = \delta_i^j$. Let~$\xi_i$ be the hamiltonian vector field corresponding to $L_i^{\mathrm{cl}}$, i.e such that $ \xi_i (f) = \{ L_i^{\mathrm{cl}}, f \} $ for any function $f$. Since $L_i^{\mathrm{cl}}$ are at most quadratic, vector fields $\xi_i$ may be exponentiated to an affine action of $G$ on $k^{2n}$. 



\vspace{1em}

\textit{Characteristic variety}

\vspace{1em}
 
Consider the variety $\Sigma = \{ (x,y) \in k^{2n} |   L_i^{\mathrm{cl}}(x,y)=0, \ i=1,...,n \}$ and its Zariski open subset $\Sigma_{s} = \{ (x,y) \in k^{2n} | dL_1^{\mathrm{cl}} \wedge ... \wedge dL_n^{\mathrm{cl}} (x,y) \neq 0 \}$. By~the implicit function theorem, $\Sigma_s$ is a manifold of dimension $n$, containing $0$. Bracket relations satisfied by $L_i^{\mathrm{cl}}$ imply that $\Sigma_s$ is Lagrangian and invariant under the $G$-action. By construction, for every $p \in \Sigma_s$ the set $\{ \left. \xi_i \right|_p \}_{i=1}^n$ is a basis of the complex tangent space of $\Sigma_s$ at $p$. In particular the holomorphic tangent bundle of $\Sigma_s$ is trivial, and the canonical bundle of $\Sigma_s$ admits a natural square root $K^{\frac{1}{2}}$ (details of its construction will be given later). Moreover the orbits of the $G$-action on $\Sigma_s$ are open in $\Sigma_s$. Since $\Sigma_s$ is a semialgebraic set, it has finitely many connected components. Therefore each connected component of $\Sigma_s$ is clopen in $\Sigma_s$. Since the orbits of $G$ in $\Sigma_s$ are open in $\Sigma_s$ and disjoint, they are precisely the connected components of $\Sigma_s$. Let $\Sigma_0$ be the connected component containing $0$. Since $\Sigma_0$ has a distinguished base point, it may be identified with the coset space $\frac{G}{\Gamma}$, where $\Gamma$ is the stabiliser of $0$ in $G$. $\Gamma$ is a~discrete subgroup of $G$. $G$ itself is then a universal covering space of $\Sigma_0$. By~the implicit function theorem, in some neighbourhood of zero $x^i$ may be used as local coordinates on $\Sigma_0$. For future reference, note that the vector fields $\xi_i$ restricted to $\Sigma_0$ and expressed in local coordinates take the form
\begin{equation}
\xi_i = \frac{\partial}{\partial x^i} - B_{ij}^k x^j \frac{\partial}{\partial x^k} - C_{i}^{jk} y_j \frac{\partial}{\partial x^k}.
\end{equation}
It is also convenient to introduce the divergence of $\xi_i$: 
\begin{equation}
\mathcal L(\xi_i)( dx^1 \wedge ... \wedge dx^n) = \mathrm{div}(\xi_i) \ dx^1 \wedge ... \wedge dx^n,
\end{equation}
where $\mathcal L(\xi_i)$ is the Lie derivative with respect to $\xi_i$. This divergence is only locally defined and coordinate dependent, so it has no geometric significance.

\newpage

\textit{Hamilton-Jacobi equation}

\vspace{1em}

Choose a symplectic potential $\theta = y_i dx^i$. Since $d \theta = \omega$ and $\Sigma_0$ is Lagrangian, restriction $\theta'$ of $\theta$ to $\Sigma_0$ is a closed form. For any $p \in \Sigma_0$ let
\begin{equation}
F_0(p) = \int_0^p \theta',
\end{equation}
with the integral taken over any smooth path from $0$ to $p$. $F_0$ is holomorphic, but may be multi-valued on $\Sigma_0$. It can be regarded as a globally defined holomorphic function on the Lie group $G$. Notice that certain function on $G$ was denoted with the same symbol as the formal series appearing in (\ref{eq:F_hbar_expansion}). This will be justified once it is shown that the two objects agree when the global version is expressed in local coordinates.

Now restrict attention to a neighbourhood $U$ of zero in $\Sigma_0$ such that $\det \left( \frac{\partial L_i}{ \partial y_j} \right) \neq 0$ on $\overline U$ and $H^1_{\mathrm{dR}}(U)=0$. A convenient choice is 
\begin{equation}
U = \{ (x,y) \in \Sigma_0 | \  |x^i| < r, \ i=1,...,n \}
\end{equation}
with some sufficiently small $r>0$. Then $\{ x^i \}$ furnish a coordinate system on $U$ and $\left. F_0 \right|_U$ is single-valued. On this coordinate patch $F_0$ may be regarded as a function of $x$, given by power series convergent uniformly on $\overline U$. Since $\theta'$ vanishes at zero, $F_0$ and $dF_0$ vanish at zero. Additionally one has $dF_0 = \theta'$, which shows that $U \cap \Sigma_0$ is a graph $y_i = \frac{\partial F_0(x)}{\partial x^i}$. Since $\left. L_i^{cl} \right|_{\Sigma_0}=0$, this implies that $F_0$ satisfies the Hamilton-Jacobi equation
\begin{equation}
L_i^{\mathrm{cl}}(x, dF_0) = \partial_i F_0 - \frac{1}{2} A_{ijk} x^j x^k - B_{ij}^k x^j \partial_k F_0 - \frac{1}{2} C_i^{jk} \partial_j F_0 \partial_k F_0= 0.
\end{equation}
Evaluating the derivative of this equation at zero one gets that the second derivatives of $F_0$ vanish at zero. Hence $F_0(x)$ satisfies the same differential equation and the same boundary condition as the unique formal series $F_0(x)$ in (\ref{eq:F_hbar_expansion}). Thus the two objects are equal. In particular $F_0(x)$ in (\ref{eq:F_hbar_expansion}) is~a~convergent series on $\overline U$. 

\vspace{1em}

\textit{Equations for the partition function}

\vspace{1em}

Equations for higher order terms of the partition function may be obtained by plugging (\ref{eq:Z_our_ansatz}) into the equation (\ref{eq:Z_annihilated}). Using the form of $\xi_i$ and $\mathrm{div}(\xi_i)$, the Hamilton-Jacobi equation satisfied by $F_0$ and the local parametrization of $\Sigma$ as the graph $y_i = \frac{\partial F_0}{\partial x^i}$ resulting system may be rewritten as
\begin{subequations}
\begin{gather}
\xi_i \rho_0 + \left( \frac{1}{2} \mathrm{div}(\xi_i) - \delta_i \right) \rho_0 =0, \label{eq:rho0} \\
\xi_i \psi_{m} = \Delta_i \psi_{m-1}, \quad \mathrm{for} \ m \geq 1, \label{eq:psi_equations}
\end{gather}
\label{eq:Zq_equations}
\end{subequations}
where $\psi_0=1$ and $\Delta_i$ are differential operators\footnote{This means that the derivatives act not only on $\rho_0^{-1}$, but also on everything to the right of $\Delta_i$.} given by
\begin{equation}
\Delta_i = \frac{1}{2} C_i^{jk} \rho_0^{-1} \partial_j \partial_k \rho_0.
\end{equation}

\vspace{1em}

\textit{Half-form $\rho_0$}

\vspace{1em}

The first step of the analysis of (\ref{eq:Zq_equations}) is to show that equation (\ref{eq:rho0}) may be solved for $\rho_0$ - holomorphic function on $U$. Let $\lambda_0$ be the holomorphic form on $U$ such that
\begin{equation}
\iota(\xi_i) \lambda_0 = \delta_i - \frac{1}{2} \mathrm{div}(\xi_i).
\label{eq:eta0}
\end{equation}
Since $\xi_i$ span every tangent space of $\Sigma_0$, this determines $\lambda_0$ uniquely. Commutation relations of $L_i$ imply a relation
\begin{equation}
\xi_i (\mu_j) - \xi_j \left( \mu_i \right) = f_{ij}^k \mu_k,
\end{equation}
where $\mu_i$ is the right hand side of (\ref{eq:eta0}). Thus $\lambda_0$ is a closed form, so $\rho_0(x)$ satisfying equation (\ref{eq:rho0}) may be written as
\begin{equation}
\rho_0(x) = \exp \left( \int_0^x \lambda_0 \right).
\end{equation}
Since $\rho_0(x)=e^{F_1(x)}$ (because the two objects are unique solutions of the same differential equation within the class of formal series satisfying certain boundary condition), series $F_1(x)$ converges in $U$.

From the geometric perspective, it is more natural to regard $\rho_0$ as a local expression for a certain half-form on $\Sigma_0$. Indeed, let $\eta$ be a section over $U$ (determined up to sign) of $K^{\frac{1}{2}}$ such that
\begin{equation}
\eta \otimes \eta = \rho_0(x)^2 \ dx^1 \wedge ... \wedge dx^n.
\end{equation}
Then equation (\ref{eq:rho0}) may be reformulated as
\begin{equation}
\mathcal L(\xi_i) \eta = \delta_i \eta.
\label{eq:eta_equation}
\end{equation}
This has the advantage that it is formulated in terms of intrinsic, coordinate independent objects on $\Sigma_0$. In particular it makes sense to ask if there exists a holomorphic section over whole $\Sigma_0$ satisfying (\ref{eq:eta_equation}). 

Let $\epsilon$ be the section of the canonical bundle of $\Sigma_0$ determined by
\begin{equation}
\epsilon \left( \xi_1, ..., \xi_n \right) =1.
\end{equation}
Simple calculation shows that $\epsilon$ is holomorphic and satisfies the equation
\begin{equation}
\mathcal L(\xi_i) \epsilon = - f_{ik}^k \epsilon.
\end{equation}
The bundle $K^{\frac{1}{2}}$ is defined as the trivial bundle with formal symbol $\sqrt{\epsilon}$ playing the role of a global framing. Tensor product $K^{\frac{1}{2}} \otimes K^{\frac{1}{2}}$ is identified with the canonical bundle $K$ of $\Sigma_0$ by the formula $f \sqrt{\epsilon} \otimes g \sqrt{\epsilon} = fg \ \epsilon$. Lie derivative of a section of $K^{\frac{1}{2}}$ with respect to a holomorphic vector field $\xi$ is defined as $\mathcal L(\xi) (f \sqrt{\epsilon}) = \xi(f) \sqrt{\epsilon} + \frac{1}{2} \frac{\mathcal L(\xi) \epsilon}{\epsilon} \sqrt{\epsilon}$, where $\frac{\mathcal L(\xi) \epsilon}{\epsilon}$ is the unique holomorphic function such that $\mathcal L(\xi) \epsilon = \frac{\mathcal L(\xi) \epsilon}{\epsilon} \epsilon$. This definition is the only one consistent with the identification $K^{\frac{1}{2}} \otimes K^{\frac{1}{2}} = K$ and the assumption that the Lie derivative satisfies the Leibniz rule.

Let $\eta = \widetilde \rho_0 \sqrt{\epsilon}$, where $\widetilde \rho_0$ is a holomorphic function. Plugging this ansatz into (\ref{eq:eta_equation}) one obtains
\begin{equation}
\xi_i (\widetilde \rho_0) = \left( \delta_i - \frac{1}{2} f_{ik}^k \right) \widetilde \rho_0.
\label{eq:tilded_rho0_equation}
\end{equation}
Commutation relations of $L_i$ and the Jacobi identity imply that $\nu_i = \delta_i - \frac{1}{2} f_{ik}^k$ satisfies $f_{jk}^i \nu_i =0$, so by the Lie's theory it coincides with the derivative at zero of a unique holomorphic homomorphism $\widetilde \rho_0$ from $G$ to the multiplicative group $\mathbb C^{\times}$ of nonzero complex numbers. By construction, $\widetilde \rho_0$ satisfies (\ref{eq:tilded_rho0_equation}). In general $\widetilde \rho_0$ is multi-valued on $\Sigma_0$. In this case global solution of (\ref{eq:eta_equation}) doesn't exist on $\Sigma_0$. Precise condition for $\widetilde \rho_0$ to descend to a function on $\Sigma_0$ is that the group $\Gamma$ should be contained in the kernel of $\widetilde \rho_0$.

If the Lie algebra $\mathfrak g$ spanned by $L_i$ coincides with its commutator ideal, then $\delta_i = f_{ik}^k=0$. Thus globally holomorphic $\eta$ on $\Sigma_0$ is guaranteed to exist.

\vspace{1em}

\textit{Quantum corrections}

\vspace{1em}

The next step is to consider equation (\ref{eq:psi_equations}). To this end, first note that commutation relations of $L_i$ imply that $\Delta_i$ commute ($[\Delta_i, \Delta_j]=0$) and satify the cocycle condtion (\ref{eq:Delta_cocycle}).

To solve (\ref{eq:psi_equations}), suppose that $\psi_{m-1}$ is some holomorphic function on $U$. Define a holomorphic $1$-form $\lambda_m$ on $U$ by
\begin{equation}
\iota(\xi_i) \lambda_m = \Delta_i \psi_{m-1}.
\label{eq:lambdan_form}
\end{equation}
Exterior derivative of $\lambda_m$ may be calculated with the aid of (\ref{eq:Delta_cocycle}):
\begin{equation}
d \lambda_m (\xi_i, \xi_j) = (\Delta_j \xi_i - \Delta_i \xi_j) \psi_{m-1}.
\end{equation}
If $\psi_{m-1}$ itself satisfies (\ref{eq:psi_equations}) with some $\psi_{m-2}$ on the right hand side, then $(\Delta_j \xi_i - \Delta_i \xi_j) \psi_{m-1} = [\Delta_j , \Delta_i] \psi_{m-2}=0$, so $\lambda_m$ is closed. Thus one may put
\begin{equation}
\psi_m(x) = \int_0^x \lambda_m.
\end{equation}
Since each $F_g$, $g \geq 2$ is a polynomial in $\psi_m$, all $F_g(x)$ are holomorphic on $U$.

In contrast to equations considered earlier in this work, (\ref{eq:psi_equations}) is not formulated in a way intrinsic to the manifold $\Sigma_0$. This is because it is not clear how to extend $\Delta_i$ to a differential operator on whole $\Sigma_0$. For this reason functions $\psi_m$ are defined only locally. 

\vspace{1em}

\textit{Structure of $\Delta$}

\vspace{1em}

Equation (\ref{eq:Delta_cocycle}) has the form of a Lie algebra cocycle condition. In particular it is automatically satisfied for $\Delta_i$ of the form $\Delta_i=[\xi_i, \kappa]$ for some second order differential operator $\kappa$. It will be shown that such $\kappa$ may always be found. Under additional assumption that $(\kappa f)(0)=0$ for any function $f$, $\kappa$~becomes uniquely determined.

Recall that on a complex Lie group there are two distinguished framings of the holomorphic tangent bundle: left-invariant (say, $\xi_i$) and right-invariant (say, $\chi_i$) vector fields, which generate right and left translations, respectively. Left-invariant fields commute with right-invariant ones.

Since $U$ is isomorphic (as a $\mathfrak g$-manifold) to a neighbourhood of the neutral element in $G$, there exists a holomorphic trivialization $\chi_i$ of $TU$ such that 
\begin{equation}
[\xi_i, \chi_j]=0, \quad \quad [\chi_i, \chi_j] = f_{ij}^k \chi_k. \label{eq:chi_relations}
\end{equation}
It is convenient to introduce also covectors $\sigma^i$ defined by $\sigma^i(\xi_j) = \delta^i_j$.

Operators $\Delta_i$ may be expanded as
\begin{equation}
\Delta_i = \frac{1}{2} \alpha_i^{pq} \xi_p \xi_q + \beta_i^p \xi_p + \gamma_i
\end{equation}
for some uniquely determined coefficient functions $\alpha, \beta, \gamma$ with $\alpha_i^{pq}=\alpha_i^{qp}$. Cocycle condition (\ref{eq:Delta_cocycle}) is equivalent to the statement that differential forms
\begin{equation}
\alpha^{pq} = \alpha^{qp}_i \sigma^i, \quad \beta^p = \beta_i^p \sigma^i, \quad \gamma = \gamma_i \sigma^i
\end{equation}
are closed. Thus there exist functions $a^{pq}, b^p$ and $c$ on $U$ such that 
\begin{equation}
\alpha^{pq} = da^{pq}, \quad \beta^p = db^p, \quad \gamma = d c. \label{eq:abc_functions}
\end{equation}
They contain $\frac{(n+1)(n+2)}{2}$ arbitrary integration constants, which may be fixed uniquely by requiring that 
\begin{equation}
a^{pq}(0)=b^p(0)=c(0)=0. \label{eq:abc_boundary_condition}
\end{equation}

Now define a differential operator
\begin{equation}
\kappa = \frac{1}{2} a^{pq} \chi_p \chi_q + b^p \chi_p + c. 
\label{eq:kappa_formula}
\end{equation}
Using (\ref{eq:chi_relations}) one gets $[\xi_i, \kappa]=\Delta_i$, which completes the proof of existence. Boundary conditions (\ref{eq:abc_boundary_condition}) are equivalent to the constraint $(\kappa f)(0)=0$ for any function $f$. To get the uniqueness statement, notice that functions $a^{pq}, b^p, c$ in (\ref{eq:kappa_formula}) are determined uniquely up to integration constants by the cocycle condition (\ref{eq:Delta_cocycle}).

Incidentally, function $c$ in (\ref{eq:kappa_formula}) may be related directly to the partition function. Indeed, the defining condition of $\kappa$ together with $\xi_i(1)=0$ implies $\xi_i(\psi_1-c)=0$. Thus assuming $c(0)=0$ one gets 
\begin{equation}
c = \psi_1.
\end{equation} 



\section{Examples} \label{sec:examples}

\subsection{One-dimensional example}

Consider the Airy structure given by a single differential operator
\begin{equation}
L = \hbar \partial_x - \frac{1}{2} x^2 - \frac{\hbar^2}{2} \partial_x^2.
\end{equation}
Replacing derivatives by $y$ and solving a quadratic equation, one finds that $\Sigma_0$ is locally given by $y(x) = 1 - \sqrt{1-x^2}$. Integrating the form $\theta = y dx$ gives
\begin{equation}
F_0(x) = x - \frac{1}{2} x \sqrt{1-x^2} - \frac{1}{2} \arcsin x.
\end{equation}
When $F_0(x)$ is expressed in terms of the global coordinate $z= \arcsin(x) \in \frac{\mathbb C}{2 \pi \mathbb Z}$ on $\Sigma_0$, it becomes clear that $F_0$ is not single valued on $\Sigma_0$. However it becomes globally holomorphic after passing to a universal cover.

Hamiltonian vector field on $\Sigma_0$ takes the form $\xi = \partial_z$. Since $\delta$ vanishes, the half-form $\eta$ is $\xi$-invariant. One may take $\eta = \sqrt{dz}$. This translates to a~local expression $\rho_0(z) = \frac{1}{\sqrt{\cos(z)}}$. A short calculation gives
\begin{equation}
\Delta =  \partial_z  \frac{1}{2 \cos^2(z)} \partial_z + \frac{3 \sin^2(z) +2}{8 \cos^4(z)}.
\end{equation}
Relation $\psi_m'(z) = \Delta \psi_{m-1}$ may be used to compute arbitrarily many $\psi_m(z)$ without expanding them in power series. For example, for $m=1$ one gets
\begin{equation}
\psi_1(z) = \frac{\sin(z) (6-\sin^2(z))}{24 \cos^3(z)}.
\end{equation}
Notice that after expressing $\psi_1$ in terms of the $x$ variable, a square root appears in the denominator. This branch-cut singularity is merely an artifact of the coordinate system. It has been checked by explicit calculation that for $m \in \{ 1,...,40 \}$ functions $\psi_m$ are meromorphic on $\Sigma_0$, with poles only on the ramification locus of the projection to the $x$ axis. It would be interesting to show that this is true for all $m$.




As promised earlier, $\Delta$ may be written as $[\xi,\kappa]$ with
\begin{equation}
\kappa = \frac{1}{4} \sin(2z) \left( \frac{1}{\cos z} \frac{\partial}{\partial z} \right)^2 + \psi_1(z). 
\end{equation}

\subsection{Borel subalgebra of $\mathfrak{sl}(2)$}

In this section two inequivalent Airy structures for a Borel subalgebra $\mathfrak b$ of~$\mathfrak{sl}(2)$ will be obtained and analyzed. Even though only generators of $\mathfrak b$ will have the form required by the definition of an Airy structure, the whole $\mathfrak{sl}(2)$ algebra is lifted to the quantum level, leading to uniqueness of the quantization. Interestingly, both Airy structures are derived from the same set of classical hamiltonians. The point is that the zero locus $\Sigma_s$ is disconnected, and each connected component gives rise to a distinct Airy structure.

The starting point is a triple of hamiltonians
\begin{subequations}
\begin{gather}
H = - e_3 e_{-3} + e_1 e_{-1}, \\
E = e_3 e_{-1} - e_1^2,  \\
F = - e_1 e_{-3} + e_{-1}^2,
\end{gather}
\end{subequations}
together with elementary Poisson brackets 
\begin{equation}
\{ e_3, e_{-3} \} = 3, \quad \{ e_1, e_{-1} \}=-1.
\end{equation}
With this convention $H,E,F$ form an $\mathfrak{sl}(2)$ triple:
\begin{equation}
\{ H, E \} = 2 E, \quad \{ H, F \} = -2 F, \quad \{ E , F \} = H. 
\end{equation}

To construct Airy structures for the Borel subalgebra $\mathfrak b$ spanned by $H$ and $E$, it is necessary to find the zero locus $\Sigma_s$ and divide it into its connected components. This is a standard computation whose details will be omitted. The result is that there are two connected components, each isomorphic to $\mathbb C \times \mathbb C^{\times}$ as a complex manifold. The first connected component $\Sigma_p$ is the orbit of the point $p:$ $e_3=1$, $e_1= e_{-1}=e_{-3}=0$, while the second component is the orbit of the point $q:$ $e_{-1}=1$, $e_3 = e_1 =e_{-3}=0$.

\vspace{1em}

\textit{The first orbit}

\vspace{1em}

To quantize $\Sigma_p$, introduce coordinates
\begin{equation}
y_1 = - e_{-3}, \quad y_2 = e_{-1}, \quad x^1 = \frac{1}{3} (e_3-1), \quad x^2 = e_1. 
\end{equation}
These satisfy standard bracket relations $\{ y_i , x^j \} = \delta_i^j$. After expressing $H,E,F$ in terms of $x$ and $y$, variables $y$ are replaced by derivatives according to the Weyl prescription. This gives
\begin{subequations}
\begin{gather}
H = \hbar(1+3 x^1) \partial_1  + \hbar x^2 \partial_2 +2 \hbar, \\
E = \hbar (1+3x^1) \partial_2 - (x^2)^2, \\
F =   \hbar x^2 \partial_1 + \hbar^2 \partial_2^2 . 
\end{gather}
\end{subequations}
Partition function may be found by solving the relevant differential equations directly. The result is
\begin{equation}
Z(x^1,x^2) = \frac{\exp \left( \frac{1}{3 \hbar} \frac{(x^2)^3}{1+3x^1} \right)}{(1+3x^1)^2}.
\end{equation}
This expression appears to have an essential singularity at $1+3x^1 =0$, but this point does not belong to $\Sigma_p$. In fact one may parametrize $\Sigma_p$ as 
\begin{equation}
e_3=e^{t}, \quad e_1=e^t s, \quad e_{-1}=e^t s^2, \quad e_{-3} = e^t s^3,
\end{equation}
with $t \in \frac{\mathbb C}{2 \pi i \mathbb Z}$, $s \in \mathbb C$. In this coordinate system $Z$ takes the form
\begin{equation}
Z(t,s) = e^{-2t} \exp \left( \frac{1}{3 \hbar} e^{2t} s \right),
\end{equation}
which is manifestly globally holomorphic.

In this example $H$ and $E$ are first order differential operators, so $\Delta$ vanishes and there are no quantum corrections at all.  

\vspace{1em}

\textit{The second orbit}

\vspace{1em}

For the second orbit, introduce coordinates 
\begin{equation}
y_1 = e_1, \quad y_2 = e_3, \quad x^1 = 1- e_{-1}, \quad x^2 = \frac{1}{3} e_{-3}.
\end{equation}
After Weyl quantization, one gets
\begin{subequations}
\begin{gather}
H = \hbar (1-x^1) \partial_1 - 3\hbar  x^2 \partial_2 - 2 \hbar  , \\
E = \hbar (1-x^1)\partial_2 -\hbar^2 \partial_1^2, \\
F = (1-x^1)^2 - 3\hbar  x^2 \partial_1.
\end{gather}
\end{subequations}
Equations $H Z = 0$, $EZ=0$ don't admit a solution which is a holomorphic function and satisfies correct boundary conditions. Indeed, if there was such a solution, it would coincide with the partition function of this Airy structure. On the other hand, it will soon be seen that this partition function is a divergent formal series.

\vspace{1em}

Hamiltonian vector fields projected to $x$ space take the form
\begin{subequations}
\begin{gather}
\xi_1 = (1-x^1) \frac{\partial}{\partial x^1} - 3 x^2 \frac{\partial}{\partial x^2}, \\
\xi_2 = (1-x^1) \frac{\partial}{\partial x^2}.
\end{gather}
\end{subequations}
Solving $(\xi_i + \frac{1}{2} \mathrm{div}(\xi_i)) \rho_0=0$ gives
\begin{equation}
\rho_0 = \frac{1}{(1-x^1)^2}.
\end{equation}
Differential operators $\Delta$ take the form
\begin{equation}
\Delta_1 = 0,  \quad \quad \Delta_2 = (1-x^1)^2 \partial_1^2 \frac{1}{(1-x^1)^2}.
\end{equation}
Therefore $\psi_m$ satisfy a hierarchy of partial differential equations,
\begin{subequations}
\begin{gather}
\left( (1-x^1) \partial_1 - 3 x^2 \partial_2 \right) \psi_m =0 , \\
(1-x^1) \partial_2 \psi_m = (1-x^1)^2 \partial_1^2 \frac{1}{(1-x^1)^2} \psi_{m-1}.
\end{gather}
\end{subequations}
The first equation states merely that each $\psi_m$ may be written in the form
\begin{equation}
\psi_m(x^1, x^2) = f_m \left( \frac{x^2}{(1-x^1)^3} \right).
\end{equation}
Plugging this into the second equation one gets
\begin{equation}
f_m(z) = \int_0^z d \zeta \ \left( 6 + 24 \zeta \frac{d}{d \zeta} + 9 \zeta^2 \frac{d^2}{d \zeta^2} \right) f_{m-1}(\zeta).
\end{equation}
This formula implies that if $f_{m-1}(z)$ is a monomial of degree $d$, then $f_m(z)$ is a monomial of degree $d+1$. Since $f_0(z)=1$, it follows that $f_m(z) = c_m z^m$, for some coefficients $c_m$. Plugging this ansatz into the equation above one gets a recurrence equation, which is easily solved: $c_m = 3(3m-1) c_{m-1}$.
Thus the partition function may be written as
\begin{equation}
Z(x^1,x^2) = \frac{1}{(1-x^1)^2} \sum_{n=0}^{\infty} \frac{\Gamma \left( n+ \frac{2}{3} \right)}{\Gamma \left(  \frac{2}{3} \right)} \left( \frac{9 \hbar x^2}{(1-x^1)^3} \right)^n.
\label{eq:divergent_Z}
\end{equation}
Proceeding as in the case of the first orbit, one may show that each term of this series is a holomorphic function on whole $\Sigma_q$. On the other hand, the series in $\hbar$ is divergent. 

\vspace{1em}

Possibility of interpreting the divergent series $\psi(w) = \sum_{m=0}^{\infty} \frac{\Gamma \left( m+ \frac{2}{3} \right)}{\Gamma \left(  \frac{2}{3} \right)} w^m$ in (\ref{eq:divergent_Z}) as an asymptotic expansion of some genuine function $Z^{\mathrm{res}}$ satisfying equations $L_i Z^{\mathrm{res}}=0$ will now be discussed. First notice that $\psi$ satisfies a~differential equation
\begin{equation}
w^2 \psi'(w) + \left( \frac{2}{3}w -1 \right) \psi(w) =-1.
\label{eq:resummation_ODE}
\end{equation}
This condition has a unique solution in the class of functions holomorphic on the pointed plane $\mathbb C^{\times}$,
\begin{equation}
\psi^{\mathrm{res}}(w) = \frac{3 e^{- \frac{1}{w} }}{w} \pFq{1}{1}{\frac{1}{3}}{\frac{4}{3}}{\frac{1}{w}},
\end{equation}
where $\pFq{1}{1}{a}{b}{z}$ is the confluent hypergeometric function. $\psi^{\mathrm{res}}(w)$ has an~essential singularity at $w=0$. By the well-known results \cite{Abramowitz} on the asymptotics of hypergeometric functions, $\psi(w)$ is an asymptotic expansion of $\psi^{\mathrm{res}}(w)$ for $w \to 0$, up to a correction term of the form $\psi^{\mathrm{cut}}(w)=\Gamma \left( \frac{1}{3} \right) e^{\frac{i \pi}{3}} e^{- \frac{1}{w}} \left( \frac{1}{w} \right)^{\frac{2}{3}}$. This term is negligibly small for $w \to 0$, provided that the real part of $w$ is positive. $\psi^{\mathrm{cut}}(w)$ is annihilated by the differential operator $w^2 \frac{d}{dw} + \left( \frac{2}{3}w -1 \right) $, so~$ \psi^{\mathrm{res}}(w)- \psi^{\mathrm{cut}}(w)$ satisfies (\ref{eq:resummation_ODE}). This function is approximated by the formal series $\psi(w)$ on a larger set than $\psi^{\mathrm{res}}(w)$, but it has branch cuts. 


Simple calculation shows that the differential equation (\ref{eq:resummation_ODE}) satisfied by $\psi^{\mathrm{res}}(w)$ is a sufficient condition for the function 
\begin{equation}
Z^{\mathrm{res}}(x^1,x^2) = \frac{1}{(1-x^1)^2} \psi^{\mathrm{res}} \left( \frac{x^2}{(1-x^1)^3} \right)
\label{eq:resummed_Z}
\end{equation}
to be annihilated by the operators $H$ and $E$. Its solutions form a one parameter family. On~the other hand, for $EZ^{\mathrm{res}}=HZ^{\mathrm{res}}=0$ to hold it is not necessary to have (\ref{eq:resummation_ODE}). A necessary and sufficient condition takes the form
\begin{equation}
3 w^2 \psi''(w) + (8w -3) \psi'(w) + 2 \psi(w)=0.
\label{eq:resummation_second_order_ODE}
\end{equation}
This equation is equal to the derivative of (\ref{eq:resummation_ODE}). Its general solution takes the form $c_1 \psi^{\mathrm{res}}(w) + c_2 \psi^{\mathrm{cut}}(w)$. This function has correct asymptotic expansion only for $c_1=1$, which is equivalent to the equation (\ref{eq:resummation_ODE}).

It may be shown that $Z^{\mathrm{res}}$ extends to a holomorphic function on the Zariski open subset of $\Sigma_0$ of all points such that $x^2 \neq 0$.


 
 \vspace{1em}

As in the previous examples, the cocycle $\Delta$ may be trivialized explicitly:
\begin{equation}
\kappa = \rho_0^{-1} \left( f \partial_1^2 - f^2 \partial_1 \partial_2 + \frac{1}{3} f^3 \partial_2^2 \right) \rho_0,
\end{equation}
where $f(x^1,x^2)=\frac{x^2}{1-x^1}$.

\subsection{Cocycle $\Delta$ not derived from an Airy structure}

Let $G$ be a connected Lie group, $\xi_i$ - basis of its Lie algebra. Pick a smooth function $f$ on $G$ vanishing at the neutral element. Define $\Delta_i = \xi_i(f)$. Since $\Delta_i$ are functions, they commute. By construction, the cocycle condition (\ref{eq:Delta_cocycle}) holds. Partition function annihilated by $T_i = \xi_i - \hbar \Delta_i$ is trivial to find:
\begin{equation}
\psi = e^{\hbar f}.
\end{equation}
Since $\Delta_i$ contains no derivatives, it can't be derived from an Airy structure.

It would be interesting to further constrain these commutative cocycles\footnote{Or perhaps more appropriately, their germs at the neutral element.} $\Delta$ on $G$ which may be obtained from some Airy structure. 

\section{Summary and outlook}

Analytic properties of the free energies associated to finite-dimensional Airy structures were investigated. It has been shown that each term of the semiclassical expansion is a convergent, rather than merely formal power series. The first two terms were exhibited as objects defined globally on a universal cover of the characteristic variety. In specific examples the same was possible also for higher order terms, up to existence of poles on the ramification locus of the projection from $\Sigma_0$ to the $x$ subspace. It is left for future works to decide if such extension is always guaranteed to exist.

For the purpose of discussion of quantum corrections as globally defined objects, it would be useful to settle the question whether it is possible to extend $\Delta_i$ to differential operators acting on, say, sections of some bundle over $\Sigma_0$. It is the author's impression that the cocycle condition discussed in the text may play a role here. It was proven that this cocycle may always be trivialized in regions with vanishing first Betti number.

Study of the equation $[[\xi_i,\kappa],[\xi_j,\kappa]]=0$ with $\kappa$ playing the role of the indeterminate is an interesting problem left for future study. The reason for   this is twofold. Firstly, it may allow to constrain Airy structures for some classes of Lie groups. Secondly, it may be hoped that $\kappa$ can be reinterpreted as a data of a quantum deformation of a Lie group.

\section*{Acknowledgments}

The author would like to thank L. Hadasz for helpful discussions during the preparation of the manuscript. This work is supported in part by the NCN grant: UMO-2016/21/B/ST2/01492.

\end{document}